\documentclass[aip,amsmath,amssymb]{4-1}

\usepackage{graphicx}
\usepackage{dcolumn}
\usepackage{bm}
\usepackage{multirow}
\usepackage{csquotes}
\usepackage{amsmath}
\begin{document}

\title{Thermal noise limited higher-order mode locking of a reference cavity}

\affiliation{MOE Key Laboratory of Fundamental Physical Quantities Measurement \\
Hubei Key Laboratory of Gravitation and Quantum Physics \\
School of Physics, Huazhong University of Science and Technology, Wuhan 430074, P. R. China}

\author{X. Y. Zeng, Y. X. Ye, X. H. Shi, Z. Y. Wang, K. Deng, J. Zhang, and Z. H. Lu}

\email{jie.zhang@mail.hust.edu.cn, zehuanglu@mail.hust.edu.cn}

\date{\today}

\begin{abstract}
Higher-order mode locking has been proposed to reduce the thermal noise limit of reference cavities. By locking a laser to the HG$_{02}$ mode of a 10-cm long all ULE cavity, and measure its performance with the three-cornered-hat method among three independently stabilized lasers, we demonstrate a thermal noise limited performance of a fractional frequency instability of $4.9\times10^{-16}$. The results match the theoretical models with higher-order optical modes. The achieved laser instability improves the all ULE short cavity results to a new low level.
\end{abstract}


\maketitle
\section{Introduction}
Ultra-stable lasers are indispensable for many experiments in optical frequency standards \cite{Ludlow:15}, gravitational wave detection \cite{Granata:10}, fundamental physics tests \cite{Oskay:05}, space applications \cite{Argence:12}, and coherent optical links \cite{Grosche:09}. A common method to achieve ultra-stable lasers is to lock free running lasers to ultra-stable high-finesse Fabry-Perot (FP) reference cavities using the Pound-Drever-Hall (PDH) technique \cite{PDH:83}.

It has been revealed that the fundamental stability limitation of a reference cavity comes from the Brownian motions of the reference cavity materials \cite{Numata:04}. Through careful design and painstaking control, ultra-stable lasers with thermal noise limited stabilities have been achieved in several labs, especially for widely used 10-cm long reference cavities made of the ultra-low expansion (ULE) material \cite{Ludlow:07, Gill:08, Hansch:08, Zhao:09, Wu:16}. In order to further reduce the thermal noise of FP cavities, one approach is to reduce the cavity temperature with cryogenic techniques, using single-crystal cavity materials with large mechanical $Q$ factors \cite{Hagemann:14, Wiens:14, Zhang:17}. Another approach is to use cavities with longer length \cite{Hafner:15}, or using lower loss angle mirror substrates and coating materials \cite{Cole:13}. The third approach is to work with larger optical modes by choosing the mirror radius of curvature (ROC) that produces a cavity close to instability \cite{Davila:17}, or using higher-order spatial cavity modes \cite{Granata:10, Mours:06, Noack:17, Notcutt:06}. The thermal noise limit of the cavity can be greatly reduced in the first two methods, but with associated higher expense and more technical problems such as vibration noise.

As for the third approach, optical modes higher than the fundamental mode have a widely spread intensity distribution, and they offer a large cancellation over the mirror substrate and coating components. It is a more economic and straightforward method, but the report of thermal noise limited utra-stable lasers based on higher-order modes is very few. In the road-map of the gravitational wave detection, the Laguerre-Gaussian 33 (LG$_{33}$) mode has long been proposed to reduce the effect of the thermal noise limit, which is one of the limiting noise sources in the current generation detector \cite{Granata:10, Mours:06, Noack:17}, but so far no experimental results concerning thermal noise limited performance are reported. To reveal fundamental thermal noise-related length fluctuations, Notcutt et al. compares the frequency instability of a laser locked to the TEM$_{00}$ mode and the TEM$_{24}$ mode of a cavity \cite{Notcutt:06}. The measured lowest frequency instability when the laser is locked to the TEM$_{24}$ mode is two times larger than the calculated thermal noise limit at a level of $1\times10^{-14}$.

In this Letter, we report a direct comparison of the frequency instability of an ultra-stable laser system that is locked to different spatial modes of a 10-cm long FP reference cavity with thermal noise limited performance. As a tradeoff, the Hermite-Gaussian 02 (HG$_{02}$) mode is chosen as the higher-order optical mode. The individual laser frequency instabilities are obtained by performing a three-cornered-hat (TCH) comparison with two other ultra-stable lasers. Thermal noise limited performances are both achieved at the HG$_{00}$ mode and the HG$_{02}$ mode, clearly matching the theoretical model prediction. A modified Allan deviation of $4.9\times10^{-16}$ is obtained when the laser is locked to the HG$_{02}$ mode. To our knowledge, this is the best result reported in an all ULE 10-cm long cavity system, demonstrating the great potential of higher-order mode locking.

\section{Thermal Noise limit for higher order cavity mode}
The thermal noise limit of reference cavities can be calculated by the fluctuation-dissipation theorem (FDT). FDT is a unique way to obtain the thermal fluctuation spectrum \cite{Levin:98, Numata:04}. Generally, the power spectral density of the cavity length displacement can be calculated as 
\begin{equation}
\label{PSD2}
S_x(f)=\frac{4k_BT}{\pi{f}}\phi{U},
\end{equation}
where $k_B$ is the Boltzmann constant, $T$ is the temperature, $f$ is the Fourier frequency, $\phi$ is the loss angle of the spacer, mirror substrate and coating, and $U$ is the strain energy stored in an FP cavity with a static pressure distribution normalized to 1 N, which is related to the laser beam radius. The expression of the parameter $U$ for the spacer, the substrate and the coating of the mirror, corresponding to the intensity distribution of the incident beam are generalized as follows \cite{Numata:04,Vinet:09,Vinet:10},

\begin{equation}
\label{Usp}
U_{nm}^{sp}=\frac{L}{6\pi R_s^2 \textrm{Y}},
\end{equation}

\begin{equation}
\label{Usb}
U_{nm}^{sb}=\frac{1-\sigma^2}{2\sqrt{\pi} \textrm{Y}\textrm{w}}\textrm{g}_{n,m}^{sb},
\end{equation}

\begin{equation}
\label{Uct}
U_{nm}^{ct}=\frac{d}{\pi \textrm{Y}\textrm{w}^2}\frac{(1-\sigma^2)(1-2\sigma)}{1-\sigma}\textrm{g}_{n,m}^{ct},
\end{equation}
where $U_{nm}^{sp}$, $U_{nm}^{sb}$, and $U_{nm}^{ct}$ are strain energies stored in the spacer, substrates and coating materials, respectively, $L$ is the length of the cavity, $R_s$ is the radius of the spacer, $d$ is the coating thickness, $\sigma$ is the Possion's ratio, $\textrm{Y}$ is the Young's modulus, $\textrm w$ is the radius of the laser beam on cavity mirrors ($\textrm w_0$ on the plane mirror and $\textrm w_1$ on the curved mirror), $\textrm{g}_{n,m}^{sb}$ and $\textrm{g}_{n,m}^{ct}$ are the $\textrm{g}$ factors of the substrates and the coatings, which depend on the distribution of the strain energy under various HG$_{nm}$ transverse mode \cite{Vinet:10}. Substituting Eq.(\ref{Usp})-Eq.(\ref{Uct}) back to Eq.(\ref{PSD2}), the frequency instability expressing as the Allan deviation is
\begin{equation}
\label{Allan}
\sigma_{y}=\frac{\alpha}{L} \sqrt{2\textrm{ln}2(S_x^{sp}(f)+2S_x^{sb}(f)+2S_x^{ct}(f))f}.
\end{equation}
Here, $\alpha =1$ for the Allan deviation and $\alpha \approx 0.82$ for the modified Allan deviation \cite{Kessler:12}. The higher the transverse optical mode number, the smaller $\textrm{g}$ factors, and therefore the smaller thermal noise limit. 

In order to realize a higher order mode locking, one can generate a higher-order mode laser beam by using spatial light modulator, phase mask \cite{Carbone:13}, or pre-locking cavity. In this work, we directly couple the HG$_{00}$ Gaussian beam into the FP cavity to get a higher-order mode by using an abnormally incident laser beam \cite{Chu:12}, with tilting and off-axis injection. To achieve a thermal noise limited locking, a high coupling efficiency is crucial in order to obtain a high signal to noise ratio and a large PDH frequency discriminator slope \cite{Zhang:16}. We numerically calculate the coupling efficiency from an HG$_{00}$ mode to the higher-order HG$_{nm}$ modes. The coupling efficiency can be calculated by the correlation function of the HG$_{00}$ mode and the HG$_{nm}$ mode \cite{Hagemann:13}:

\begin{equation}
\label{couple}
\eta_{nm}= \frac{|\iint^{\infty}_{-\infty}E_{00}(x,y)E_{nm}^{*}(x,y)\textrm{d}x\textrm{d}y|^2 }{\iint^{\infty}_{-\infty}|E_{00}(x,y)|^2\textrm{d}x\textrm{d}y\iint^{\infty}_{-\infty}|E_{nm}|^2\textrm{d}x\textrm{d}y},
\end{equation}
where the $E_{00}(x,y)$ and $E_{nm}(x,y)$ are the electric field distributions of the injected HG$_{00}$ mode and the coupled higher-order HG$_{nm}$ mode, respectively. 

\begin{figure}[htbp]
    \centering
    \includegraphics[width=\linewidth]{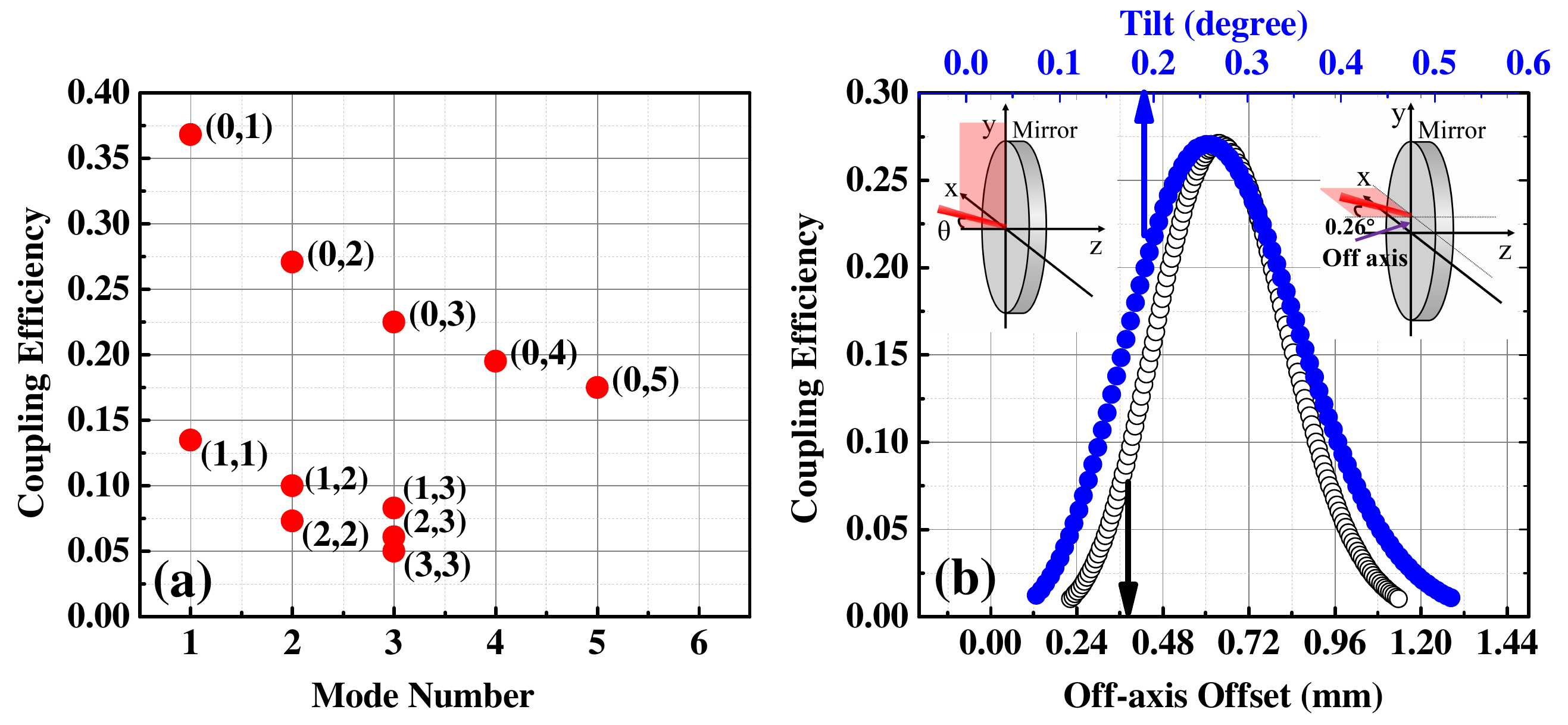}
    \caption{(a)Maximum coupling efficiency from an HG$_{00}$ mode incident laser beam to an HG$_{nm}$ cavity mode. (b) Simulation of the coupling efficiency of the HG$_{02}$ mode. Insets: Schemes of the incident laser beam relative to the plano mirror. The gray disk stands for the coating layer of the mirror, the red beam is for the incident laser beam, and the transparent red area is the plane formed by the incident laser beam and the $z$ axis.}
    \label{coup}
\end{figure}

Figure \ref{coup} (a) presents the numerical results of the maximum coupling efficiency from the HG$_{00}$ incident laser beam to an HG$_{nm}$ cavity mode by changing the incident tilt angle and the off-axis offset. The maximum coupling efficiency is $\eta_{01}= 36.7\%$, $\eta_{02}= 27.1\%$ and $\eta_{03}= 22.5\%$ for unsymmetric modes, and $\eta_{11}=13.5\%$, $\eta_{22}=7.3\%$ for symmetric modes. It is quite hard to get a high coupling efficiency with symmetric modes or other composite modes by direct coupling from an HG$_{00}$ input mode. Figure \ref{coup} (b) shows the simulation of the coupling efficency of the HG$_{02}$ mode, considering both offset and tilt angle of the incident laser beam. As shown in the inset of the Fig. \ref{coup} (b), the coupling efficiency reaches maximum either with a  0.26$^\circ$ tilt angle, or with a 0.63 mm offset at the $y$ axis under a 0.26$^\circ$ tilt angle at the $x-z$ plane. As a tradeoff, we choose the HG$_{02}$ mode for a demonstration of higher-order mode locking. According to Eqs.(1-5), the modified Allan deviations of the thermal noise limit for the HG$_{00}$ mode and the HG$_{02}$ mode are $5.9\times10^{-16}$ and $4.8\times10^{-16}$, respectively. There is a $18\%$ reduction of the thermal noise limit from HG$_{00}$ mode to HG$_{02}$ mode. ($L=0.1$ m, $R_s= 0.035$ m, $d=5$ $\mu$m, $\sigma=0.18$, $\textrm{Y}$=67.9 Gpa, $\textrm w_0$=261 $\mu$m, $\textrm w_1$=291 $\mu$m and the scaling factors $\textrm{g}_{0,2}^{sb}=0.683$ and $\textrm{g}_{0,2}^{ct}=0.641$ are used for the calculation \cite{Vinet:10}.)

\begin{figure}[htbp]
    \centering
    \includegraphics[width=\linewidth]{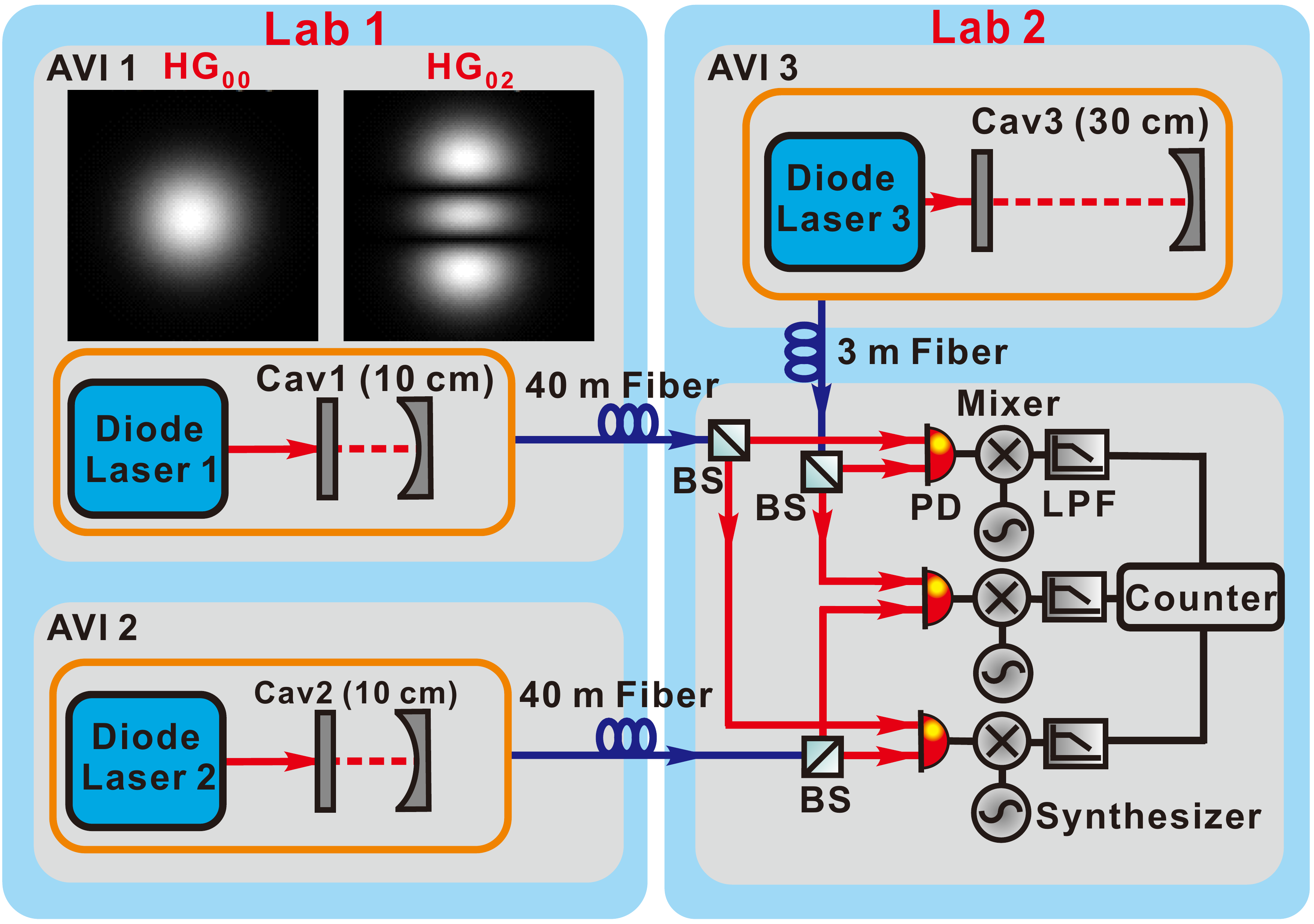}
    \caption{Experimental scheme of the TCH measurement. AVI, active vibration isolation table; BS, beam splitter; PD, InGaAs photo diode; LPF, low pass filter.}
    \label{setup}
\end{figure}

To verify this thermal noise limit reduction, we lock a diode laser to the HG$_{00}$ mode and the HG$_{02}$ mode of a 10-cm long ultra-stable cavity, separately. To characterize the achieved frequency instability, we beat the laser frequency with two other independent laser systems through the rigorous TCH method, considering the correlations of these three lasers \cite{Premoli:93}. Figure \ref{setup} shows the TCH measurement scheme. All the lasers operate at a wavelength of 1070 nm, whose 4th harmonic wavelength at 267.4 nm can be used for the Al$^+$ ions clock transition. The two 10-cm long all ULE cavity systems that located at lab1 are described in detail in Ref \cite{Zhang:16}. The two cavities are designated as Cav1 and Cav2, respectively. The frequency instability of the system using Cav1 is evaluated and compared when the laser is locked to the HG$_{00}$ mode and the HG$_{02}$ mode of this cavity. The other two laser systems are stabilized to their fundental modes as references. The finesses of the HG$_{00}$ mode and the HG$_{02}$ mode for Cav1 are about $1.3\times10^{5}$ and $3.2\times10^{5}$ respectively, measured by the cavity ring down technique. The finesses of the HG$_{00}$ mode of Cav2 is around $3.3\times10^{5}$. Cav3 located at lab2 is a 30-cm long ULE cavity with fused silica mirrors and ULE rings. The finesse of Cav3 for the HG$_{00}$ mode is about $3.8\times10^{5}$. 

The output of the three ultrastable lasers are sent to a beat detection unit in lab2 through polarization-maintaining fibers. With active fiber noise cancellation, the residual fiber noise contribution is in an order of $1\times10^{-17}$ at 1 s, which is negligible for the laser frequency instability evaluation. The beat signals of the three lasers under different cavity modes are all within 600 MHz, detected independently with three InGaAs photodiodes. They are first mixed down to 1.8 MHz, and filtered by low pass filters with a bandwidth of 1.9 MHz. We record the three beat signals with a high resolution multichannel synchronous phase recorder (K+K Messtechnik, model FXE65) in the phase averaging mode with a 100 ms gate time.

To reach a thermal noise limited locking performance for Cav1, we take great cares on the system design and environment control \cite{Zhang:16, Zhang2:16}. To reduce the temperature fluctuation of the cavity, Cav1 is housed in a vcauum chameber with a gold-plated copper shield layer as a low-pass thermal filter. The pressure of the vacuum chamber is evacuated to a level of  $1\times10^{-6}$ Pa, and the thermal time constant from the vacuum chamber to the ULE cavity is measured to be about one day. Besides, the temperature of the vacuum chamber is stabilized with a digital controll loop at the zero crossing temperature of Cav1, which is T$_{0}$=36.8 ℃ and the slope of the thermal expansion coefficience of Cav1 at this temperature is measured to be $1.2\times10^{-9}$ /K$^2$. The temperature fluctuation is about 1 mk during 24 hours. In order to isolate the acoustic noise, temperature and pressure fluctuation, we place the whole system including the vacuum chamber and the optical setup for the PDH locking on an active vibration isolation (AVI) table and enclose it in a box made by stainbless steel plates, covered with acoustic absorption foams. The hardwares of Cav2 are almost the same with Cav1, and they are independently placed on two AVI tables, as shown in Fig.~\ref{setup}. 

In order to reduce the residual amplitude modulation (RAM) effect of the electro-optic phase modulator (EOM) in the PDH locking, we place ioslators with 35 dB isolation in the front and the back of the EOM. Furthermore, the crystal of the EOM is temperature controlled to a level of fluctuation less than 10 mK. The evaluated RAM to frequency instability for Cav1 system is in the order of $1\times10^{-16}$ at an averaging time of less than 10 s. The incident laser beam matches the HG$_{00}$ mode of Cav1 with a 261 $\mu$m radius, shaped with a lens pair. In order to reach a thermal noise limited performance, we use an incident optical power of 50 $\mu$W into the photodiode for the PDH error signal detection, and the error signal is amplified to increase the piezo feedback gain at the low frequency range. The PDH locking bandwidth through a fast current feedback branch is around 2 MHz, and the locking bandwidth of a slow piezo feedback loop is around 5 kHz. For the higher-order mode locking, we adjust the coupling mirror pair in the front of Cav1 to tilt the incident laser beam into the HG$_{02}$ mode. The obtained coupling efficiency is 24\%, which is close to the theoretical calculation 27\%.

\begin{figure}[htbp]
\centering
\includegraphics[width=\linewidth]{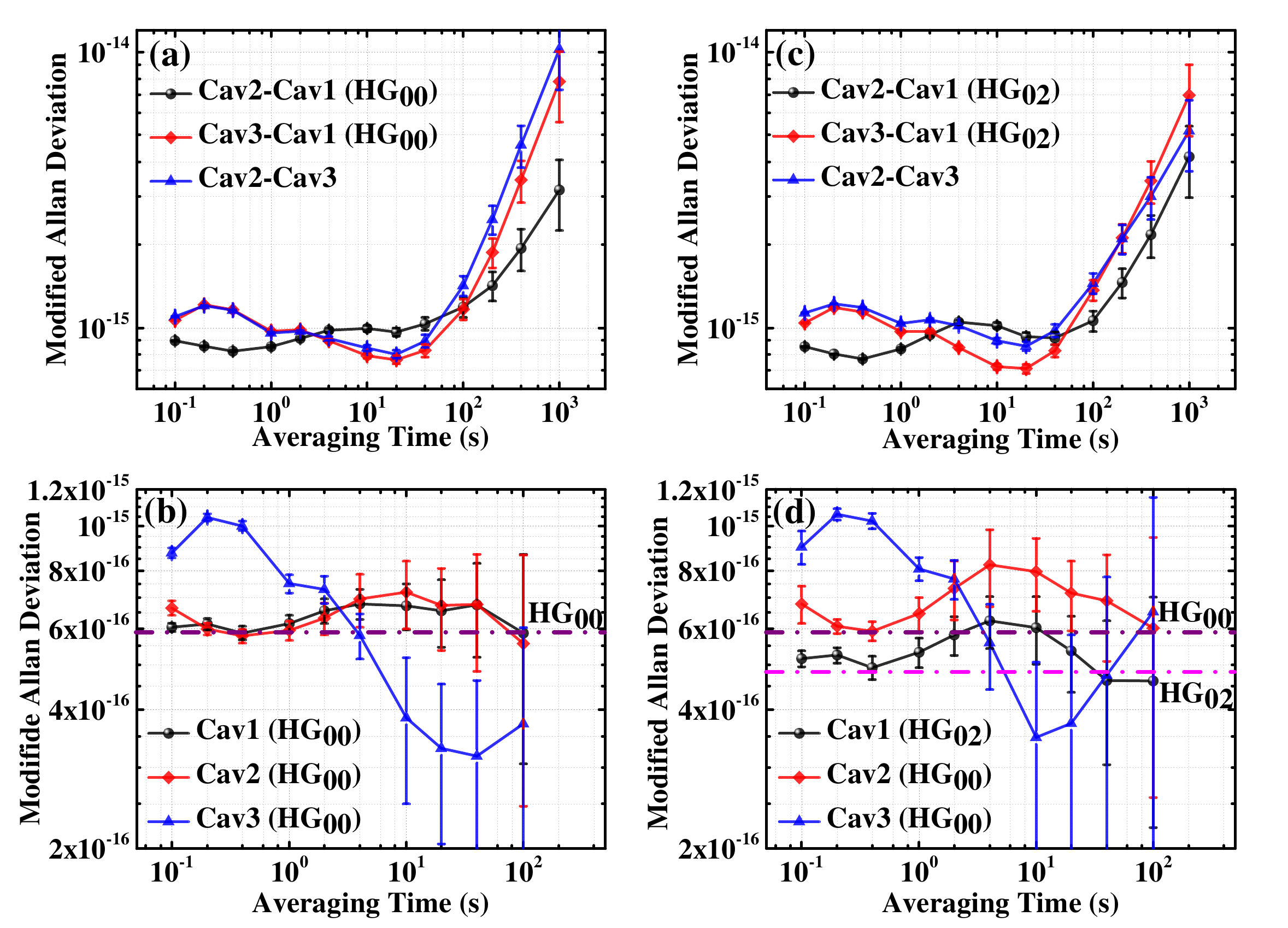}
\caption{(a) Modified Allan deviation of frequency instability of three beat signals when diode laser 1 is locked to the HG$_{00}$ mode of Cav1. We remove linear drifts for the 10 hours data. Cav1-Cav2: 5 mHz/s; Cav1-Cav3: -78 mHz/s; Cav2-Cav3: 83 mHz/s; (b) Modified Allan deviation of the three individual lasers when diode laser 1 is locked to the HG$_{00}$ mode of Cav1. (c) Modified Allan deviation of frequency instability of three beat signals when diode laser 1 is locked to the HG$_{02}$ mode of Cav1. We remove linear drifts for the 10 hours data. Cav1-Cav2: -9 mHz/s; Cav1-Cav3: 40 mHz/s; Cav2-Cav3: 31 mHz/s; (d) Modified Allan deviation of the three individual lasers when diode laser 1 is locked to the HG$_{02}$ mode of Cav1.}
\label{Modified Allan deviation}
\end{figure}

The TCH measurement results are shown in Fig.~\ref{Modified Allan deviation}. Figure~\ref{Modified Allan deviation} (a) and (b) show the results when the diode laser 1 is locked to the fundamental mode (HG$_{00}$) of Cav1. Figure~\ref{Modified Allan deviation} (a) shows the modified Allan deviation of the measured beat frequencies using a 10 hours continuously recorded data. The frequency instability of the beat frequency between Cav1 and Cav2 reaches $8.2\times10^{-16}$ at 0.4 s. Figure~\ref{Modified Allan deviation} (b) shows the frequency instability of three individual lasers. The 10 hours data is split into 60 data sets with a duration of 600 s. We remove linear drifts and perform the TCH analysis for each data set. The error bars stand for the statistic standard deviation of the 60 data sets \cite{Kessler:12}. The frequency instabilities for both Cav1 and Cav2 are closed to $5.9\times10^{-16}$ at 0.1 s to 1 s, limited by the thermal noise of the HG$_{00}$ mode, shown with the purple dash-dot line. 

Figure~\ref{Modified Allan deviation} (c) and (d) show the results when diode laser 1 is locked to the HG$_{02}$ mode of Cav1. As shown in Fig.~\ref{Modified Allan deviation} (c), the beat frequency instability between Cav1 and Cav2 with a 10 hours continuously data reaches $7.7\times10^{-16}$ at 0.4 s, lower than the case of the HG$_{00}$ mode locking. Figure~\ref{Modified Allan deviation} (d) shows the TCH analysis. The frequency instabilities of Cav2 and Cav3 stay at the same level, but the instabilities of Cav1 clearly decrease and reach $4.9\times10^{-16}$ at 0.4 s, which is very colse to the calculated thermal noise limit of the HG$_{02}$ mode shown with the pink dash-dot line. For both locking cases, the correlations between these laser systems are evaluated. The averaged modified Allan covariances between these three lasers are all around $1\times10^{-46}$ from 0.1 s to 4 s, showing minimum correlations at this time scale. The averaged modified Allan covariances from 10 s to 100 s are around$4\times10^{-34}-8\times10^{-32}$, suggesting possible temperature correlation effect \cite{Premoli:93}.

\begin{figure}[htbp]
\centering
\includegraphics[width=\linewidth]{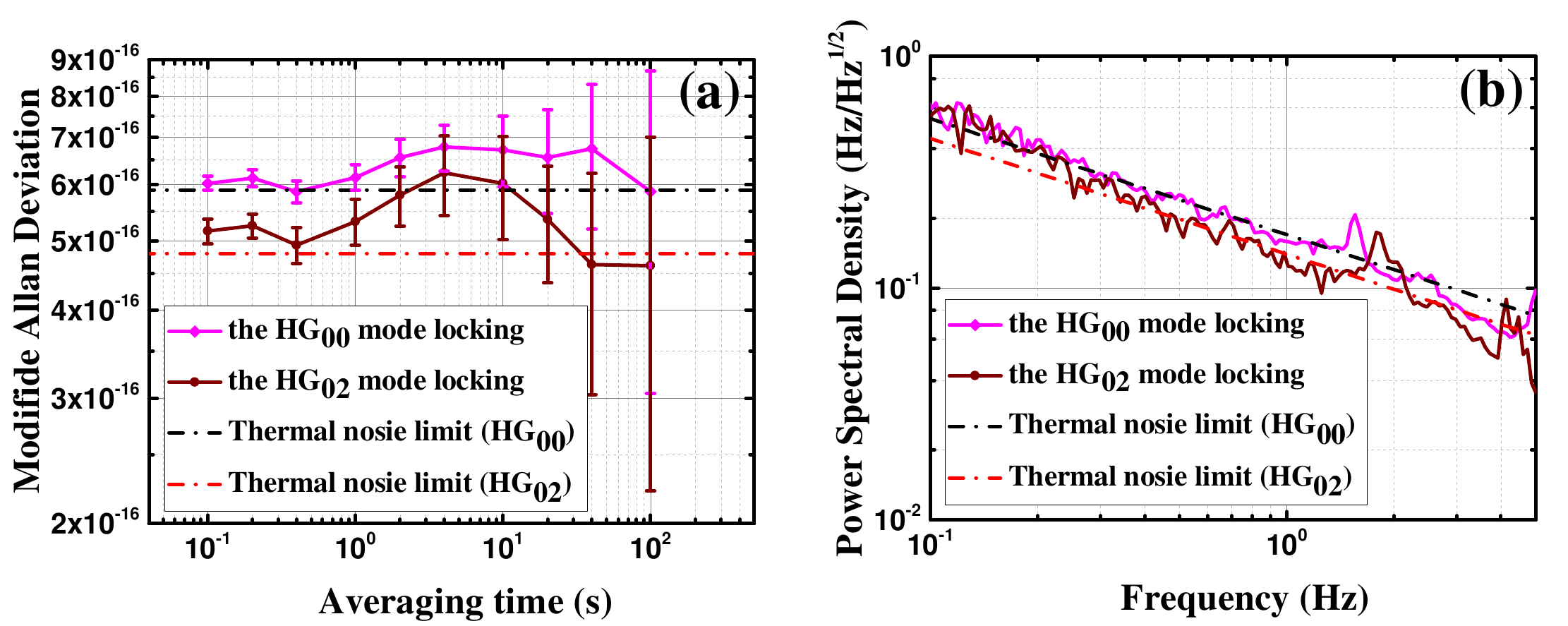}
\caption{(a) Modified Allan deviations of diode laser 1 when it is locked to the HG$_{00}$ mode and the HG$_{02}$ mode of Cav1, respectively. (b) Frequency noise PSD of diode laser 1 in the two locking cases. The dash dot lines show the theoretical thermal noise limit of HG$_{00}$ mode and the HG$_{02}$ mode, which are 0.17/$\sqrt{f}$  $\mathrm{Hz/\sqrt{Hz}}$ and 0.14/$\sqrt{f}$ $\mathrm{Hz/\sqrt{Hz}}$, respectively.}
\label{frequency PSD}
\end{figure}

For clarity, we redraw the frequency instabilities of diode laser 1 when it is locked to the HG$_{00}$ mode and the HG$_{02}$ mode of Cav1 in Fig.~\ref{frequency PSD} (a). In both cases, the thermal noise limited locking are achieved, and agree well with the theroetical thermal noise limit. The frequency instability of the laser reaches $4.9\times10^{-16}$ with the higher-order mode locking. To our knowledge, this is the best result achieved among all similar designed 10-cm long all ULE cavities \cite{Ludlow:07, Hansch:08, Gill:08, Zhao:09, Wu:16}. It is even comparable with the results of 10-12 cm long ULE cavities with fused silica mirrors \cite{Santarelli:09, Keller:14}. We also calculate the phase noise power spectral density (PSD) in these two cases using cross-correlation spectrum method \cite{Xie:17}, and then covert them into frequency noise PSD, as shown in Fig.~\ref{frequency PSD} (b). The peaks around 1.5 Hz may be caused by the resonant frequency of AVI1 where Cav1 sits. The frequency noise PSDs for the HG$_{00}$ mode locking and the HG$_{02}$ mode locking are in good agreement with the theoretical thermal noise frequency PSDs for nearly two decades frequency range.

In conclusion, we realize thermal noise limited locking using both a fundamental mode and a higher-order mode, demonstrating the potential of thermal noise limit reduction to $10^{-16}$ level by using higher-order mode locking. We obtain a frequency instability of $4.9\times10^{-16}$, to our knowledge this is the best result among all similar designed 10-cm long all ULE cavities. Higher order mode locking is a very promising way for further improvement of the ultrastable clock lasers' performances, since it is already the limitation of many optical clocks. In addition, taking into account cavity length and compactness, higher-order mode locking has advantages in portable systems and space applications. For instance, for a 10-cm long ULE cavity with a 10 m ROC concave fused silica mirror pair \cite{Davila:17}, the thermal noise limit of the HG$_{55}$ mode will be as low as $5.9\times10^{-17}$ in modified Allan deviation. In the future, we plan to implement a spatial light modulator into our system so that we can have higher coupling efficiency for the higher-order modes with even smaller thermal noise limit.

The project is partially supported by the National Key R\&D Program of China (Grant No. 2017YFA0304400), the National Natural Science Foundation of China (Grant Number 91536116, 91336213, and 11774108).

\end{document}